\documentclass[%
 reprint,
 amsmath,amssymb,
 aps,
]{revtex4-2}

\usepackage{graphicx}% Include figure files
\usepackage{dcolumn}% Align table columns on decimal point
\usepackage{bm}% bold math
\begin{document}

\preprint{APS/123-QED}

\title{Quantum Brachistochrone for the Majorana Particle}% Force line breaks with \\
\thanks{Thanks to UTS MAPS for supporting this research}%

\author{Peter G. Morrison}
 \email{nanoscope@outlook.com}
\affiliation{%
 Department of Maths and Physical Sciences, University of Technology, Sydney.\\}%

\date{\today}% It is always \today, today,

\begin{abstract}
This paper covers some new results from the theory of time optimal quantum control, with particular application to relativistic particles including Majorana fermions. We give a brief review of the state of affairs regarding experimental results, and a concise overview of the methodology of time optimal control and unitary transformation. This technique is then applied to the Majorana particle and used to derive four dimensional conservation laws. We then discuss the nature of time transformation in these systems and some simple scattering problems. 

\end{abstract}

\maketitle

%\tableofcontents

\section{\label{sec:level1}Introduction}
One of the outstanding problems in modern quantum electrodynamics is the nature of the Majorana particle. From the direction of the Dirac theory of electron spin, there are no clear physical reasons as to why this particle should exist or not. As such we are required to find ways in which these systems may be distinguished from one another. If there is a physical reason for a particle's non-existence, this reason should be clear and unambiguous. We shall show in this paper how one may easily distinguish one from the other, and a possible reason for the difficulties in producing these exotic types of particles.

This paper shall briefly run over the available experimental data and tests that are currently being carried out in this field to detect this theoretical construct for completeness. As we will be working with systems over the 4x4 matrices, we shall also cover a number of other simple physical situations that exist on this space. For example, as we shall show, it is a simple exercise to set up a problem such that the Hamiltonian matrix encodes information for the 4D angular momentum, via the Runge-Lenz vector. It is a matter of physical consequence that such a quantity is naturally conserved.

The method of the quantum brachistochrone provides clear insight into the nature of fundamental symmetry within many different physical systems. Given that previous papers have brought the concepts of electromagnetic interaction via spin, strong nuclear forces and relativistic electrons \cite{morrison2008time1, morrison2010time1, morrison2012time2, morrison2019time1, morrison2019time2} under the same umbrella, it is hoped that this further demonstration of the utility of this technique will bring it into focus. Originally developed by Carlini et. al
\cite{carlini2005quantum, carlini2006time, carlini2007time, carlini2008time, carlini2011time, carlini2012time, carlini2014time} the area of time optimal control is enjoying a period of rapid growth. However, many examples that have been calculated thus far focus almost solely on spin chain systems of one sort or another. This paper shall use a different method, in that we shall be using spinors, which are related to representations of relativistic particles which have anticommutation relations instead of commutation laws.

\section{\label{sec:level1}Review}
We shall give an overview of the main topics in this paper in sequential order. Many of the results contained in this paper are potentially testable in an experimental context, so a concise rundown of the groups that are working in similar fields in the laboratory will be given where appropriate. The topic addressed is broad, and there is significant overlap in the results from many different areas, which we shall comment on wherever it applies.
\subsection{\label{sec:level2}Quantum Brachistochrone}

The quantum brachistochrone, originally addressed in a series of papers by Carlini et. al
\cite{carlini2005quantum,carlini2006time, carlini2007time} and followed closely by the works of the author, see e.g. \cite{morrison2008time1} is a sophisticated technique of open loop control of quantum systems, that seeks to exploit the underlying periodic behaviour of the quantum state to understand optimal trajectories from one quantum state to another. The classical brachistochrone is familiar as the curve that takes least time given constraints; such problems of analysis are familiar since Bernoulli and possibly quite longer. The quantum analog of this problem focuses on the application of either variational calculus or unitary operations to achieve optimal orbits on the space of states. This topic is recently enjoying a number of successes, and it appears to be a generally applicable and somewhat sensible theory for understanding complicated quantum systems. We note the series of works in \cite{morrison2010time1, morrison2012time2, morrison2019time1, morrison2019time2} move away from the standard examples of spin systems, to examine time optimal control for qutrit systems related to nuclear quadrupole resonance, and higher order systems related to the Dirac equation and spinor states. Other results in spin systems and Ising chains may be found in \cite{carlini2011time}, transfer of coherence in \cite{carlini2012time} and Gaussian states related to quantum optics in \cite{carlini2014time}.

\subsection{\label{sec:level2}Majorana Particle}
This paper shall address some ongoing concerns with the detection of the Majorana particle, a proposed fundamental state of nature which has never been observed. The papers of Ettore Majorana \cite{majorana1932atomi, majorana1937teoria, majorana2006symmetric} c. 1932-1937, see \cite{aguado2017majorana} for translated version, address the fundamental question of symmetry between electron and positron. As is well known now- but less so in 1932, the particles we observe in nature have corresponding antiparticles with properties that are associated with one another. For example, electrons and positrons have identical mass, but opposing charges of equal strength. The particle hypothesised through symmetry arguments of Majorana has the interesting and intriguing property that in a sense it is its own antiparticle. The contrast in predicted behaviour between a Dirac system and that of Majorana has led many to speculate on the reasons why electrons and positrons are easy to produce and observe, but exotic chiral states as in a Majorana particle have never been observed in magnitude, if at all.

There is a particular industrial need for this question to be resolved in one way or another; various groups have proposed the use of Majorana particles as exotic topological quantum states within quantum computing devices, see for example \cite{aasen2016milestones, leijnse2012introduction, beenakker2013search}. However, most recently, there has been progress in a negative direction, with experiments failing to confirm the existence of said quantum states, most recently reported in \cite{chang2021absence}, also \cite{frolov2021quantum}. This leads one to question whether there is a missing piece of information in understanding the particular difficulty in producing these types of quantum states. As we shall show, there is a link between the nature of these types of systems and the underlying invariance principles as expressed through the theory of time optimal quantum state control. 

\subsection{\label{sec:level2}4D Angular Momentum}
The use of 4-vectors for describing properties of a relativistic system is well known, but this paper shall use an alternative formalism. We shall be using the angular momentum tensor, which encodes all the properties of angular momentum transformations in a manifestly invariant way. This is not the primary focus of the paper, however we do direct the reader to several resources that are of utility in this calculation. Note that the angular momentum tensor is generated using a bivector approach, which distinguishes it from the cross-product spinor calculation normally used in quantum mechanics.

For a concise overview of the application of the 4-D angular momentum principle with applications to kinematics, a historical reference is given by \cite{moses1965kinematics}, also later consult \cite{yamaleev1984quantum} for a discussion of the application of angular momentum within 4-D Euclidean space, also see \cite{chiu2017angular} for an extensive overview of the use of the Poincare group in quantum field theory. Reference texts addressing the topic may be found e.g. see \cite{shapiro2017group} or \cite{ydri2019modern}. 

\section{\label{sec:level1}Theory of Time Optimal Control}
We shall give a brief overview of the method of time optimal quantum control. This method, while closely allied to the concept of the quantum brachistochrone, is more general than the method of Carlini et al. \cite{carlini2005quantum}, as we are not bound to reference target and initial states of the system. For the most part, the methodology is similar, but there are significant differences when it comes to evaluation of the unitary operator. We shall be using the method developed in \cite{morrison2019time1, morrison2019time2}, in particular we direct the reader to the second paper as it is most closely aligned with the analysis in this document.

Time optimal quantum control is open loop control of quantum systems in order to push the state from one configuration to another in least time. The objective function to be optimised in this case is the time taken from any accessible point to another on the projective manifold which describes the space of states. Classically, the brachistochrone is well known since the times of J. Bernoulli and I. Newton, and forms the basis for many other physical ideas, including Huyghen's principle and Fermat's law of ray optics. The action principle we shall take will have the following properties: it shall minimise the time taken, it will constrain the total energy to some finite maximum, and it will allow us to choose the accessible degrees of freedom available to the system. The last property can be understood in terms of control fields, in the simplest situation. We can apply a magnetic field on some co-ordinates to force the system to be in some particular set of configurations. For example, in a non-relativistic spin-1/2 system, we may apply a magnetic field in some direction of our choosing, relative to the fixed axis of the laboratory, e.g. if we take
\begin{equation}
\mathbf{B}=B_{x}(t)\mathbf{e}_{x}    
\end{equation}
the Hamiltonian matrix may be written as

\begin{equation}
\tilde{H}=\mathbf{B}\cdot\hat{\mathbf{\sigma}}=B_{x}(t)\hat{\sigma}_{x}    
\end{equation}
In this simple situation, it is easy to see that the constraint, which we represent by the matrix \begin{math}\hat{F}\end{math} is a linear combination of the components in the matrix vector space which do not appear in the Hamiltonian, i.e.
\begin{equation}
\tilde{F}=\lambda_{y}(t)\hat{\sigma}_{y}+\lambda_{z}(t)\hat{\sigma}_{z}
\end{equation}
as by the basic properties of the Lie algebra we have 
\begin{equation}
\mathrm{Tr}\left[\tilde{H}\tilde{F}\right]=0  
\end{equation}
which represents the concept of linear independence in this vector system. Mathematically, therefore, the action principle may be written in the form:
\begin{equation}
    S=\int1dt+\mathrm{constraints}
\end{equation}
where we follow the initial assumptions of Carlini et. al \cite{carlini2005quantum} in setting up the problem.The first part of the action principle is equivalent to the time used in the Fermat action principle to calculate paths of light rates in a medium with a variable refractive index. In our situation, as our states are evolving on the projective manifold defined by the space of \begin{math}\mathbb{CP}^{n-1}\end{math}. This allows us to use the Fubini-Study metric in order to analyse the situation. The following property is easily shown using the basic formulae of quantum mechanical expectation and statistics:
\begin{equation}
    \Delta E^{2}=\left\langle \Psi\right|\tilde{H}^{2}(t)\left|\Psi\right\rangle -\left(\left\langle \Psi\right|\tilde{H}(t)\left|\Psi\right\rangle \right)^{2}
\end{equation}
 We shall discuss briefly the nature of the constraints; we can propose the following principles. Firstly, the state of the system evolves unitarily via a Schrödinger equation, we have \begin{math}
i\left|d\Psi\right\rangle =i.d\left|\Psi\right\rangle =\tilde{H}(t)\left|\Psi\right\rangle dt
\end{math}, in which case we find:
\begin{equation}
1dt=\dfrac{1}{\Delta E}\sqrt{\left\langle d\Psi\right|\left(1-\hat{P}\right)\left|d\Psi\right\rangle }    
\end{equation}
 where \begin{math}
\hat{P}=\left|\Psi\right\rangle \left\langle \Psi\right|
\end{math}
is the projection operator onto the state 
\begin{math}\Psi
\end{math}
. This contributes to the Lagrangian a constraint of the form:
\begin{equation}
    i\left\langle \dot{\phi}\left.\right.\psi\right\rangle +\left\langle \phi\left.\tilde{H}\right.\psi\right\rangle +\mathrm{c.c.}
\end{equation}
where \begin{math}
\left|\phi\right\rangle 
\end{math} is a Lagrange multiplier. For the full details of the calculation, consult \cite{morrison2008time1, carlini2005quantum}. We shall not persue the full method of Euler-Lagrange equations, rather, we shall take the basic underlying quantum evolution as given and use some simple considerations to arrive at the same result. Given that we assume that the system evolves unitarily, the von Neumann equation reads as:
\begin{equation}
    i\dfrac{d\hat{A}}{dt}=[\tilde{H}(t),\hat{A}]
\end{equation}

for any observable operator given by a Hermitian matrix, 
\begin{math}
\hat{A}^{\dagger}=\hat{A}
\end{math}
. Now, by definition, the energy eigenstates are generated by an observable, we must therefore have 
\begin{math}
\tilde{H}^{\dagger}=\tilde{H}
\end{math}
, and by extension the constraint is also Hermitian. The operator given by \begin{math}
\hat{A}=\tilde{H}+\tilde{F}
\end{math} is hence a Hermitian operator, substituting this into the equation above, we find the matrix differential equation:
\begin{equation}
    i\dfrac{d}{dt}\left(\tilde{H}(t)+\tilde{F}(t)\right)=[\tilde{H}(t),\tilde{F}(t)]
\end{equation}

which we term the quantum brachistochrone equation.
The second and third parts of the constraint may be understood in the following way. We require that the energy of the state be some finite value, this implies that the variance is finite:
\begin{equation}
    \Delta E=\sqrt{\left\langle \tilde{H}^{2}\right\rangle -\left\langle \tilde{H}\right\rangle ^{2}}
\end{equation}
It is also possible to show through the consideration of gauge invariance that the Hamiltonian is trace-free:
\begin{equation}
    \mathrm{Tr}\left[\tilde{H}(t)\right]=0
\end{equation}
This implies that we have the following expression which we term the isotropic constraint:
\begin{equation}\mathrm{Tr}\left[\dfrac{\tilde{H}^{2}}{2}\right]-k=0
\end{equation}
The final piece of information we require in order to determine the Hamiltonian matrix is that it be linearly independent of the constraint, via the trace relation:
and \begin{equation}
\mathrm{Tr}\left[\tilde{H}\tilde{F}\right]=0
\end{equation}
this is sufficient to specify the behaviour of the Hamiltonian, and thence give the dynamics of the system as a whole. This can be understood in terms of an analogy between the dot product between two orthogonal vectors, which is also zero.
As we shall show, there are various other techniques that we can use to evaluate the quantum state. The time evolution operator is defined through the formula 
\begin{equation}
\hat{U}(t,s)\left|\Psi(s)\right\rangle =\left|\Psi(t)\right\rangle 
\end{equation} or alternatively via the integral given by:
\begin{equation}
    \hat{U}(t,s)=\exp\left(-i\int_{s}^{t}\tilde{H}(\tau)d\tau\right)
\end{equation}
To be completely correct, this should be considered as a time-ordered exponential, however for the systems in this paper we can rely on some properties of the Hamiltonian to simplify the evaluation of this complicated function. We may write the Hamiltonian in a diagonal reference frame as:
\begin{equation}
    \tilde{H}(t)=\hat{W}(t)\hat{D}\hat{W}^{-1}(t)
\end{equation}
Now, we know that the Hamiltonian evolves unitarily via
\begin{equation}
    \tilde{H}(t)=\hat{U}(t,s)\tilde{H}(s)\hat{U}^{\dagger}(t,s)
\end{equation}. Calculating this using the formula above, it is simple to show that we have the matrix equation:
\begin{equation}
    \hat{U}(t,s)=\hat{W}(t)\hat{W}^{-1}(s)
\end{equation}
For certain physical systems, it may be that the matrix operator \begin{math}\hat{W}(t)\end{math} may have symmetries, but we shall not assume such to maintain complete generality at this stage. The physics is clear, we are using a time dependent transformation to diagonalise a time dependent operator. From the basis of time dependent eigenstates, the unitary operator is composed of one branch propagating forwards in time and one in reverse. For many examples that have been calculated, time translation invariance is available and a natural consequence of this method of calculation. We shall not assume this either, for completeness. These are the basic formulae within this schemata that we shall use to analyse various quantum systems in this paper. We note the contrast with the method in \cite{carlini2005quantum}; we are able to evaluate the unitary operator, time dependent state and every property of the system without the difficulty of evaluating boundary conditions. This method is designed with unitary evolution in mind; there are complexities associated with the systems we expect to encounter through use of gamma matrices and spinors. However, as demonstrated in \cite{morrison2019time1,morrison2019time2} it is possible to work quite consistently in higher dimensional spaces using this method.

\section{\label{sec:level1}Majorana Particle}
We shall briefly discuss the setup of the problem. Majorana particles \cite{majorana1932atomi,majorana1937teoria, majorana2006symmetric} have the peculiar property that, under the appropriate transformations as given by the Dirac theory of the electron, they transform as their own antiparticle. Such a feature has not been observed in nature, and the purpose of this paper is to show how the theory of the quantum brachistochrone is able to discriminate a “theoretical” particle from a real, observable particle with proper characteristics. This will be a toy problem in this sense, but in carrying out the calculation we will demonstrate a method of wide applicability to understanding quantum systems. The Dirac equation, within this formalism, is given by the fundamental matrix identity:
\begin{equation}
    \tilde{H}=m\hat{\beta}+\hat{\mathbf{\alpha}}\cdot\mathbf{p}
\end{equation}
with \begin{math}\tilde{H}^{2}=(m^{2}+|\mathbf{p}|^{2})\mathbf{1}\end{math}. Unitary evolution occurs as before, and this system has been solved completely for the quantum brachistochrone in \cite{morrison2019time1}. Now, we shall show that for a particular choice of representation for the matrices (\begin{math}\hat{\mathbf{\alpha}},\hat{\beta}\end{math}) we can recover the Majorana equation. Firstly, let us choose this representation so that all of these matrices are real. In this case we will have the time evolution equation:
\begin{equation}
    i\dfrac{d}{dt}\left|\Psi\right\rangle =(i\hat{\beta}m+\hat{\mathbf{\alpha}}\cdot\mathbf{p})\left|\Psi\right\rangle 
\end{equation}
Note that in this calculation, the momentum and mass have as yet arbitrary time dependence, to be determined from the dynamics defined by the quantum brachistochrone; we suppress the notation when convenient. Considering further, one particular representation of the Majorana matrices is given by the set defined by:
\begin{equation}
    \begin{array}{ccc}
\hat{\beta}=\left[\begin{array}{cccc}
0 & 0 & 1 & 0\\
0 & 0 & 0 & 1\\
-1 & 0 & 0 & 0\\
0 & -1 & 0 & 0
\end{array}\right] &  & \hat{\alpha}_{x}=\left[\begin{array}{cccc}
1 & 0 & 0 & 0\\
0 & -1 & 0 & 0\\
0 & 0 & -1 & 0\\
0 & 0 & 0 & 1
\end{array}\right]\\
\\
\hat{\alpha}_{y}=\left[\begin{array}{cccc}
0 & 0 & 1 & 0\\
0 & 0 & 0 & 1\\
1 & 0 & 0 & 0\\
0 & 1 & 0 & 0
\end{array}\right] &  & \hat{\alpha}_{z}=\left[\begin{array}{cccc}
0 & 1 & 0 & 0\\
1 & 0 & 0 & 0\\
0 & 0 & 0 & -1\\
0 & 0 & -1 & 0
\end{array}\right]
\end{array}
\end{equation}

Careful evaluation shows that \begin{math}\hat{\gamma}_{0}=\hat{\beta}\hat{\alpha}_{x}\hat{\alpha}_{y}\hat{\alpha}_{z}\end{math} is given by the matrix:
\begin{equation}
\hat{\gamma}_{0}=\left[\begin{array}{cccc}
0 & -1 & 0 & 0\\
1 & 0 & 0 & 0\\
0 & 0 & 0 & 1\\
0 & 0 & -1 & 0
\end{array}\right]
\end{equation}
Now, we expect that the algebra of interest will be anticommutative. Indeed, we have immediately that 
\begin{equation}
    \left\{ \hat{\gamma}_{0},\hat{\alpha}_{i}\right\} =\left\{ \hat{\gamma}_{0},\hat{\beta}\right\} =0
\end{equation} by inspection. Amongst the other operators, we have as usual:
\begin{equation}
    \left\{ \hat{\beta},\hat{\alpha}_{i}\right\} =\left\{ \hat{\alpha}_{i},\hat{\alpha}_{j}\right\} =0
\end{equation}

for \begin{math}
i\neq j
\end{math} as required. The square relationship gives \begin{equation}
    \hat{\alpha}_{i}^{2}=\mathbf{1}=\hat{\beta}^{2}
\end{equation}. The commutator brackets give further physical properties, defining spin operators via
\begin{equation}
\hat{S}_{i}=\dfrac{1}{2}\left[\hat{\alpha}_{j},\hat{\alpha}_{k}\right]    
\end{equation}
 we have e.g.
 \begin{equation}
     \begin{array}{cc}
\hat{S}_{x}=\left[\begin{array}{cccc}
0 & 0 & 0 & -1\\
0 & 0 & -1 & 0\\
0 & 1 & 0 & 0\\
1 & 0 & 0 & 0
\end{array}\right] & \hat{S}_{y}=\left[\begin{array}{cccc}
0 & 1 & 0 & 0\\
-1 & 0 & 0 & 0\\
0 & 0 & 0 & 1\\
0 & 0 & -1 & 0
\end{array}\right]\\
\end{array}
 \end{equation}
 
 \begin{equation}
     \hat{S}_{z}=\left[\begin{array}{cccc}
0 & 0 & 1 & 0\\
0 & 0 & 0 & -1\\
-1 & 0 & 0 & 0\\
0 & 1 & 0 & 0
\end{array}\right]
 \end{equation}
which close under commutation once more to give the standard law of quantum mechanics:
\begin{equation}
    \hat{S}_{i}=\dfrac{i}{2}\epsilon_{ijk}\left[\hat{S}_{j},\hat{S}_{k}\right]
\end{equation}
Thus far, this is standard quantum theory, and is completely in agreement with modern analysis of the Majorana particle. We shall now apply the methodology of the quantum brachistochrone in order to show the particular difficulty associated with this differential equation. If we take the spanning space to be the set of matrices defined by the Kronecker product, i.e. \begin{equation}
    \hat{\Upsilon}_{ij}=\hat{\sigma}_{i}\otimes\hat{\sigma}_{j}, with \hat{\sigma}_{i}\in\left\{ \hat{\sigma}_{x},\hat{\sigma}_{y},\hat{\sigma}_{z},\mathbf{1}_{2}\right\}
\end{equation} it is simple to see that we have:
\begin{equation}
    \begin{array}{c}
\hat{\beta}=-i\hat{\sigma}_{y}\otimes\mathbf{1}_{2}\\
\hat{\alpha}_{x}=\hat{\sigma}_{z}\otimes\hat{\sigma}_{z}\\
\hat{\alpha}_{y}=\hat{\sigma}_{x}\otimes\mathbf{1}_{2}\\
\hat{\alpha}_{z}=\hat{\sigma}_{y}\otimes\hat{\sigma}_{x}
\end{array}
\end{equation}
Now, by inspection we can see that the Hamiltonian as defined is not strictly Hermitian. Indeed, the adjoint equation in this instance is given by:
\begin{equation}
    \left\langle \bar{\Psi}\right|=\left\langle \Psi\right|\hat{\beta}
\end{equation}
so we must proceed carefully. It is simple to show that for this example we have the expression 
\begin{equation}
\tilde{H}^{2}=(m^{2}+|\mathbf{p}|^{2})\mathbf{1} 
\end{equation}
as we would expect; however, in order to make the Hamiltonian strictly Hermitian one must insert the imaginary unit into the equation. We shall put this to the side for now and focus on the solution of the problem. Writing out the constraint for the system, we have:
\begin{equation}
    \tilde{F}=\sum_{i,j\in H^{c}}\lambda_{ij}(t)\hat{\Upsilon}_{ij}
\end{equation}
where the subscript on the sum indicates that it is taken over all members \begin{math}\hat{\Upsilon}_{ij}\end{math} that are in the complementary set of \begin{math}\tilde{H}\end{math} as discussed earlier. We shall now demonstrate some simple tricks that aid in analysis of the quantum brachistochrone. If we evaluate the matrix operator 
\begin{math}
\tilde{F}
\end{math}
in terms of the parameters, and then compute the commutator via \begin{math}[\tilde{H},\tilde{F}]\end{math}, a simple way to find the underlying differential equations is therefore to multiply the quantum brachistochrone equation by a particular generator of the Lie algebra and then take traces. As Lie algebras are trace independent spaces, we then find, e.g.:
\begin{equation}
    \mathrm{Tr}\left[i\dfrac{d}{dt}(\tilde{H}+\tilde{F})\hat{g}\right]=\mathrm{Tr}\left[[\tilde{H},\tilde{F}]\hat{g}\right]
\end{equation}
By inserting different choices of \begin{math}\hat{g}\end{math} and relying on the concept of trace independence, it is simple to read off the differential equations. We shall not be evaluating them in depth in this paper, but note that we have:
\begin{equation}
    \mathrm{Tr}\left[[\tilde{H},\tilde{F}]\hat{\alpha}_{x}\right]=8i(\lambda_{ty}p_{z}+\lambda_{xz}m+\lambda_{yz}p_{y})=i\dfrac{dp_{x}}{dt}
\end{equation}
for example, and the other differential equations may be evaluated in a similar fashion. We can see immediately that this is a difficult way to approach this problem, as it requires us to manage a large system of simultaneous DEs which is unwieldy as the dimension increases. It is not also directly clear how to find the particular invariants; and as such, we shall use the method outlined in \cite{morrison2019time1} to solve the problem. The eigenvector matrix is readily evaluated for the Hamiltonian as being:
\begin{equation}
    \hat{W}=\left[\begin{array}{cccc}
\dfrac{p_{z}}{p_{y}-im} & \dfrac{E+p_{x}}{p_{y}-im} & \dfrac{p_{z}}{p_{y}-im} & -\dfrac{(E+p_{x})}{p_{y}-im}\\
\dfrac{E-p_{x}}{p_{y}-im} & \dfrac{p_{z}}{p_{y}-im} & -\dfrac{(E+p_{x})}{p_{y}-im} & \dfrac{p_{z}}{p_{y}-im}\\
0 & 1 & 0 & 1\\
1 & 0 & 1 & 0
\end{array}\right]
\end{equation}
with inverse:
    \begin{equation}
     \hat{W}^{-1}=\dfrac{1}{2E}\left[\begin{array}{cccc}
0 & -p_{y}+im & -p_{z} & (E+p_{x})\\
-p_{y}+im & 0 & (E-p_{x}) & -p_{z}\\
0 & -p_{y}+im & p_{z} & (E-p_{x})\\
-p_{y}+im & 0 & (E+p_{x}) & p_{z}
\end{array}\right]
\end{equation}
where we have assumed that \begin{math}
E^{2}=m^{2}+|\mathbf{p}|^{2}
\end{math}. A matter of calculation then shows that these matrices diagonalise the Hamiltonian in the sense that we have:
\begin{equation}
    \hat{W}^{-1}\tilde{H}\hat{W}=E\left[\begin{array}{cccc}
1 & 0 & 0 & 0\\
0 & 1 & 0 & 0\\
0 & 0 & -1 & 0\\
0 & 0 & 0 & -1
\end{array}\right]=\hat{D}
\end{equation}

for any values of mass and momentum. We have not specified the time dependence as yet; let us now show how this is achieved. The matrix of eigenstates evolves according to the matrix differential equation:

\begin{equation}
    i\dfrac{d\hat{W}}{dt}=\tilde{H}(t)\hat{W}(t)
\end{equation}and we also know that we will have furthermore that \begin{equation}
    \tilde{H}(t)\hat{W}(t)=\hat{D}\hat{W}(t)
\end{equation} therefore
\begin{equation}
    \hat{W}(t)=\exp\left(-it\hat{D}\right)\hat{W}(0)
\end{equation}.
This is the method of evaluation we will use. Direct substitution then gives:
\begin{widetext}
\begin{equation}
    \hat{W}(t)=\left[\begin{array}{cccc}
\dfrac{e^{-2iEt}p_{z}}{p_{y}-im} & \dfrac{e^{-2iEt}(E+p_{x})}{p_{y}-im} & \dfrac{e^{-2iEt}p_{z}}{p_{y}-im} & -\dfrac{e^{-2iEt}(E-p_{x})}{p_{y}-im}\\
\dfrac{e^{-2iEt}(E-p_{x})}{p_{y}-im} & \dfrac{e^{-2iEt}p_{z}}{p_{y}-im} & -\dfrac{e^{-2iEt}(E+p_{x})}{p_{y}-im} & \dfrac{e^{-2iEt}p_{z}}{p_{y}-im}\\
0 & 1 & 0 & 1\\
1 & 0 & 1 & 0
\end{array}\right]
\end{equation}
\begin{equation}
    \hat{W}^{-1}(t)=\dfrac{1}{2E}\left[\begin{array}{cccc}
0 & e^{2iEt}(p_{y}-im) & -p_{z} & (E+p_{x})\\
e^{2iEt}(p_{y}-im) & 0 & (E-p_{x}) & -p_{z}\\
0 & -e^{2iEt}(p_{y}-im) & p_{z} & (E-p_{x})\\
-e^{2iEt}(p_{y}-im) & 0 & (E+p_{x}) & p_{z}
\end{array}\right]
\end{equation}
Note that, as per standard results of quantum mechanics, we are free to take a universal phase factor of \begin{math}e^{-iEt}\end{math} over the whole system of eigenstates with no change to the essential physics. We are now in a position to evaluate the time evolution operator; calculating, we obtain:
\begin{equation}
    \hat{U}(t,s)=\hat{W}(t)\hat{W}^{-1}(s)=\left[\begin{array}{cccc}
e^{-2iE(t-s)} & 0 & 0 & 0\\
0 & e^{-2iE(t-s)} & 0 & 0\\
0 & 0 & 1 & 0\\
0 & 0 & 0 & 1
\end{array}\right]
\end{equation}
or \begin{equation}
\hat{U}(t,s)=e^{-iE(t-s)}\exp\left(-i(t-s)\hat{D}\right)
\end{equation}
We plainly have time translation invariance via 
\begin{math}
\hat{U}(t,s)=\hat{U}(t-s,0)
\end{math} and the standard properties of unitary inverses. Let us now apply this operator to the initial Hamiltonian to see what results, computing the linear algebra we have:
\begin{equation}
    \tilde{H}(t)=\hat{U}(t,0)\tilde{H}(0)\hat{U}^{\dagger}(t,0)
\end{equation}
\begin{equation}
\tilde{H}(t)=\left[\begin{array}{cccc}
p_{x} & p_{z} & (p_{y}+im)e^{-2iEt} & 0\\
p_{z} & -p_{x} & 0 & (p_{y}+im)e^{-2iEt}\\
(p_{y}-im)e^{2iEt} & 0 & -p_{x} & -p_{z}\\
0 & (p_{y}-im)e^{2iEt} & -p_{z} & p_{x}
\end{array}\right]
\end{equation}
with initial Hamiltonian we assumed given by:
\begin{equation}
    \tilde{H}(0)=i\hat{\beta}m_{0}+\hat{\mathbf{\alpha}}\cdot\mathbf{p}_{0}=\left[\begin{array}{cccc}
p_{x} & p_{z} & (p_{y}+im_{0}) & 0\\
p_{z} & -p_{x} & 0 & (p_{y}+im_{0})\\
(p_{y}-im_{0}) & 0 & -p_{x} & -p_{z}\\
0 & (p_{y}-im_{0}) & -p_{z} & p_{x}
\end{array}\right]
\end{equation}
\end{widetext}
ergo, we conclude that 
\begin{math}
m(t)=m_{0}e^{2iEt}
\end{math}
, similar for \begin{math}
p_{y}(t)
\end{math}, with \begin{math}
p_{x}(t)=p_{x}, p_{z}(t)=p_{z}
\end{math}. We can immediately see therefore that this equation will not transform correctly as the mass is not a constant of motion. This is to be contrasted with the analogous calculation for the Dirac equation, considered in \cite{morrison2019time2}, where it was shown that the rest mass of the electron is a constant of motion. There is obviously a relationship between these two sets of equations but we can see the essential difficulty clearly. A well defined particle will have a constant rest mass. This is necessary to ensure consistency of such expressions as conservation of momentum and energy, as a sinusoidal change in the rest mass value would have incalculable consequences for any interaction of such a hypothetical particle with its surroundings. We take the atomistic position on such things; it is clearly inconsistent with the nature we observe to have such difficulties. It is possible that the conclusion is therefore that the rest mass of such a particle must therefore be identically zero. Further discussion on the topic is required before this can be resolved, as this would create certain further problems in the nature of interaction, as such a particle would only be able to interact via angular momentum transfer.

It is apparent through this calculation that there is a requirement for an appending axiom to be added to the principles of time optimal quantum control. Although it is possible to manufacture any number of hypothetical resonant states and consequently particles using this methodology, we must apply a strict criterion in setting up the problem to model the real, observed particles that exist within the universe. That the rest mass of the particle as defined in the calculation has the correct transformation under the quantum brachistochrone to preserve its value would appear to be the minimal adjustment to the framework in order to achieve this objective. 

It is interesting to consider other representations of Dirac-type operators in this space; indeed, by using isometry, it is possible to show that the two examples of Majorana and Dirac may be transformed to any other representation on the 4x4 matrices. We will not consider the representation theory of differential operators in this paper other than a note in the conclusions, but this an area of profitable enquiry which has yet to be investigated fully. Other topics that should be computed include the nature of the relationship between the special unitary group SU(3,1), the special orthogonal group SO(3,1) and the imaginary constant in the Hamiltonian operator. It would be interesting to see whether it is possible to modify the constructions within this paper to deal with more general non-Hermitian spaces. By continuing the matrix operators to complex time in the appropriate way, one should be able to recover the hyperbolic equivalents for rotation which we associate with dilations in time.

\section{\label{sec:level1}Four Dimensional Angular Momentum}
As is well known from quantum mechanics, the four-dimensional angular momentum is describable by the Pauli-Lubanski pseudovector, viz.:
\begin{equation}
  W_{\mu}=\dfrac{1}{2}\epsilon_{\mu\nu\rho\sigma}M^{\nu\rho}P^{\sigma}  
\end{equation}
where \begin{math}
M^{\nu\rho}
\end{math} is the angular momentum tensor. It is this matrix we are interested in, it may be written as:
\begin{equation}
    M^{\nu\rho}\doteq\tilde{M}=\left[\begin{array}{cccc}
0 & -N_{x} & -N_{y} & -N_{z}\\
N_{x} & 0 & L_{xy} & -L_{zx}\\
N_{y} & -L_{xy} & 0 & L_{yz}\\
N_{z} & L_{zx} & -L_{yz} & 0
\end{array}\right]
\end{equation}
Now, by multiplying with the imaginary constant, we arrive at an operator with suitable characteristics as a Hermitian operator. We shall take the following matrix to be our toy Hamiltonian for this system:
\begin{math}
\tilde{H}=i\tilde{M}
\end{math}. We shall show a simple trick that will work with any system possessing symmetry of this type. This is obviously antisymmetric under inversion, and the constraint law may be written as:
\begin{equation}
    \tilde{F}=\left[\begin{array}{cccc}
\omega_{1} & \xi_{1} & \xi_{2} & \xi_{3}\\
\xi_{1} & \omega_{2} & \xi_{4} & \xi_{5}\\
\xi_{2} & \xi_{3} & \omega_{3} & \xi_{6}\\
\xi_{4} & \xi_{5} & \xi_{6} & \omega_{4}
\end{array}\right]
\end{equation}with 
\begin{equation}
   \omega_{1}+\omega_{2}+\omega_{3}+\omega_{4}=0 
\end{equation}
Computation of the linear algebra gives the necessary requirement of trace independence, via 
\begin{math}\mathrm{Tr}\left[\tilde{H}\tilde{F}\right]=0\end{math}.
We therefore have the quantum brachistochrone equation given by:
\begin{equation}
i\dfrac{d}{dt}\left(\tilde{H}+\tilde{F}\right)=\left[\tilde{H},\tilde{F}\right]    
\end{equation}
which reduces to 
\begin{equation}
\dfrac{d}{dt}\left(-\tilde{M}+i\tilde{F}\right)=i\left[\tilde{M},\tilde{F}\right]    
\end{equation}
and we have immediately that \begin{math}\tilde{M}(t)=\tilde{M}(0)\end{math} and
\begin{equation}
   \dfrac{d\tilde{F}}{dt}=\left[\tilde{M},\tilde{F}\right] 
\end{equation}
Evaluation of the following sets of equations produces the same result:
\begin{equation}
\mathrm{Tr}\left[i\dfrac{d}{dt}(\tilde{H}+\tilde{F})\hat{g}\right]=\mathrm{Tr}\left[\left[\tilde{H},\tilde{F}\right]\hat{g}\right]    
\end{equation}
Choosing for \begin{math}
\hat{g}
\end{math} the generators of the Hamiltonian, e.g. 
\begin{equation}
\hat{g}=\left[\begin{array}{cc}
\hat{\sigma}_{y} & \mathbf{0}\\
\mathbf{0} & \mathbf{0}
\end{array}\right]    
\end{equation}
we have: 
\begin{equation}
    2i\dfrac{dN_{x}}{dt}=0
\end{equation}
and likewise for all the other components of the Pauli-Lubanski tensor. This matrix is therefore a constant of motion, and defines conserved properties of the system, which is what we would expect. We can see that the nature of the geometry is such that:
\begin{equation}
\mathrm{Tr}\left[\dfrac{\tilde{H}^{2}}{2}\right]=2(\mathbf{L}\cdot\mathbf{L}+\mathbf{N}\cdot\mathbf{N})    
\end{equation}
To simplify the following analysis, we shall take our initial conditions such that \begin{math}
N_{y}=N_{z}=0
\end{math} In accordance with this, all angular momentum components in the yz direction must be kept, in this particular instance we have:
\begin{equation}
\tilde{H}=i\left[\begin{array}{cccc}
0 & -N_{x} & 0 & 0\\
N_{x} & 0 & 0 & 0\\
0 & 0 & 0 & -L_{yz}\\
0 & 0 & L_{yz} & 0
\end{array}\right]
\end{equation}
In the diagonal frame, we have:
\begin{equation}
\tilde{H}=\hat{W}\hat{D}\hat{W}^{-1}    
\end{equation}
as before, with the eigenmatrix
\begin{equation}
\hat{W}=\left[\begin{array}{cccc}
-i & i & 0 & 0\\
1 & 1 & 0 & 0\\
0 & 0 & -i & i\\
0 & 0 & 1 & 1
\end{array}\right]
\end{equation}
\begin{widetext}

with diagonal matrix of eigenvectors given by:
\begin{equation}
\hat{D}=\left[\begin{array}{cccc}
N_{x} & 0 & 0 & 0\\
0 & -N_{x} & 0 & 0\\
0 & 0 & L_{yz} & 0\\
0 & 0 & 0 & -L_{yz}
\end{array}\right]    
\end{equation}
Evaluating the time evolution operator in this circumstance, we find:

\begin{equation}
\hat{U}(t,0)=\hat{W}\exp\left(-it\hat{D}\right)\hat{W}^{-1} =\left[\begin{array}{cccc}
\cos(N_{x}t) & -\sin(N_{x}t) & 0 & 0\\
\sin(N_{x}t) & \cos(N_{x}t) & 0 & 0\\
0 & 0 & \cos(L_{yz}t) & -\sin(L_{yz}t)\\
0 & 0 & \sin(L_{yz}t) & \cos(L_{yz}t)
\end{array}\right] 
\end{equation}

\end{widetext}
One can see how by combining the appropriate rotations and dilations one can produce all the necessary rotations to describe the total evolution of the system. As the translational energy-momentum is decoupled from the angular momentum, which we can clearly see through the spherical relationship, we can build up the total evolution by implementing each rotation in turn without much extra complication. The formula for the total time evolution operator is lengthy and it does not add to the analysis, so for brevity we omit it.

This calculation has shown once more that the quantum brachistochrone method can be used in higher dimensional physics to derive consistent answers. All experimental evidence points towards the constancy of relativistic angular momentum. That we are able to derive this is an advance in our understanding of the underlying principles that drive this well-understood and familiar phenomenon.

\section{\label{sec:level1}Comparison between Dirac and Majorana systems}
The earlier results have indicated that the Majorana particle will not transform in the same way under the quantum brachistochrone as the Dirac type particle discovered in earlier works \cite{morrison2019time1}. One can see how at any instant of fixed time, the operator feels as if it is physically possible. This hidden symmetry between momentum and mass operators is potentially one way in which to distinguish physical from hypothetical objects. We know for a fact that Dirac electrons exist. Majorana particles “feel” like electrons. This could be one reason why their observed behaviour and difficulties in particle production differs from the reality we observe using current theory. We derived, using the quantum brachistochrone:

\begin{widetext}
\begin{equation}
\tilde{H}(t)=\left[\begin{array}{cccc}
p_{x} & p_{z} & (p_{y}+im)e^{-2iEt} & 0\\
p_{z} & -p_{x} & 0 & (p_{y}+im)e^{-2iEt}\\
(p_{y}-im)e^{2iEt} & 0 & -p_{x} & -p_{z}\\
0 & (p_{y}-im)e^{2iEt} & -p_{z} & p_{x}
\end{array}\right]    
\end{equation}
with initial Hamiltonian we assumed given by:
\begin{equation}
\tilde{H}(0)=i\hat{\beta}m_{0}+\hat{\mathbf{\alpha}}\cdot\mathbf{p}_{0}=\left[\begin{array}{cccc}
p_{x} & p_{z} & (p_{y}+im_{0}) & 0\\
p_{z} & -p_{x} & 0 & (p_{y}+im_{0})\\
(p_{y}-im_{0}) & 0 & -p_{x} & -p_{z}\\
0 & (p_{y}-im_{0}) & -p_{z} & p_{x}
\end{array}\right]    
\end{equation}

for the Majorana particle algebra, specified through the pseudo-spinor as specified. Contrasting with this, for the Dirac spinor, we have Hamiltonian matrix as derived in \cite{morrison2019time2}:
\begin{equation}
\tilde{H}(t)=\left[\begin{array}{cc}
m\mathbf{1} & -ie^{-2iEt}\mathbf{p}_{0}\cdot\mathbf{\sigma}\\
ie^{+2iEt}\mathbf{p}_{0}\cdot\mathbf{\sigma} & -m\mathbf{1}
\end{array}\right]
\end{equation}
At face value, these matrix dynamical systems do not appear to be significantly different. However, close inspection gives the crucial nature of where they part ways. 
The phase associated with the momentum term in the matrix above cancels out in any meaningful product. However, with the Majorana particle, it is not possible to separate the mass and momentum in a coherent fashion when transformed under the quantum brachistochrone. It is this essential difficulty which is obscured by the time independent perspective of quantum mechanics, which is made most clear through the analysis and methods calculated in this paper.

\end{widetext}

\section{\label{sec:level1}Scattering Theory of Majorana Particle}
We shall show some simple results now that we have derived the necessary Hamiltonian matrices. For example, other papers \cite{morrison2012time2} have shown that it is possible to use Feynman's method \cite{feynman1998quantum, feynman2018theory} to derive the Compton effect, using the spinors derived. It would be desirable to carry out a similar exercise using the pseudo-spinors that comprise the Majorana matrices as calculated earlier. Following Feynman \cite{feynman1998quantum, feynman2018theory}, we can write the scattering problem for the Compton effect as:

\begin{equation}
    m\hat{\gamma}_{t}=\mathbf{p}_{1}
\end{equation}
\begin{equation}
    E_{2}\hat{\gamma}_{t}+ip_{2}(\hat{\gamma}_{x}\cos\phi+\hat{\gamma}_{y}\sin\phi)=\mathbf{p}_{2}
\end{equation}
\begin{equation}
    \omega_{1}(\hat{\gamma}_{t}+i\hat{\gamma}_{x})=\mathbf{q}_{1}
\end{equation}
\begin{equation}
    \omega_{2}(\hat{\gamma}_{t}+i(\hat{\gamma}_{x}\cos\theta+\hat{\gamma}_{y}\sin\theta))=\mathbf{q}_{2}
\end{equation}

where our representation may be written as 
\begin{equation}
    i\mathbf{p}-m\hat{\gamma}_{t}=\mathbf{p}_{F}-m\hat{\gamma}_{t}
\end{equation} relative to Feynman's expressions \cite{feynman2018theory} as determined by the solution for the spinor brachistochrone as in \cite{morrison2019time2}. The algebra as determined in the above representation of the gamma matrices is defined by 
\begin{equation}
\{\hat{\gamma}_{t},\hat{\gamma}_{i}\}=0    
\end{equation}
\begin{equation}
  \hat{\gamma}_{\alpha}^{2}=\mathbf{1}_{4}  
\end{equation}
\begin{equation}
    \{\hat{\gamma}_{i},\hat{\gamma}_{j}\}=2\delta_{ij}
\end{equation}
\begin{equation}
    [\hat{\gamma}_{i},\hat{\gamma}_{j}]=\epsilon_{ijk}\hat{\gamma}_{k}
\end{equation}
We can show matrix identities such as:
\begin{equation}
\mathbf{q}_{1}^{\dagger}\mathbf{q}_{1}=\omega_{1}^{2}(2\mathbf{1}+i[\hat{\gamma}_{t},\hat{\gamma}_{x}])    
\end{equation}
Squaring the final momentum term, we have:

\begin{equation}
    \mathbf{p}_{2}^{2}=(\mathbf{p}_{1}+\mathbf{q}_{1}-\mathbf{q}_{2})^{2}
\end{equation}from which we obtain:
\begin{equation}
\mathbf{p}_{2}^{2}=\mathbf{p}_{1}^{2}+\left\{ (\mathbf{q}_{1}-\mathbf{q}_{2}),\mathbf{p}_{1}\right\} -\left\{ \mathbf{q}_{1},\mathbf{q}_{2}\right\}    
\end{equation}
 where 
 \begin{equation}
 \mathbf{q}_{1}^{2}=\mathbf{q}_{2}^{2}=0    
 \end{equation}
 by using the expressions for the anticommutation relations. Computing, we find:
 \begin{equation}
 \mathbf{p}_{1}^{2}=m^{2}\mathbf{1}
  \end{equation}
  \begin{equation}  
  \mathbf{p}_{2}^{2}=(E_{2}^{2}-p_{2}^{2})\mathbf{1}
  \end{equation}
  \begin{equation}  
  \left\{ (\mathbf{q}_{1}-\mathbf{q}_{2}),\mathbf{p}_{1}\right\} =2m(\omega_{1}-\omega_{2})\mathbf{1}
  \end{equation}
  \begin{equation}
  \left\{ \mathbf{q}_{1},\mathbf{q}_{2}\right\} =2\omega_{1}\omega_{2}(1-\cos\theta)\mathbf{1}
  \end{equation}
 
 and we find the result:
 \begin{equation}
 (E_{2}^{2}-p_{2}^{2}-m^{2})\mathbf{1}=(2m(\omega_{1}-\omega_{2})-2\omega_{1}\omega_{2}(1-\cos\theta))\mathbf{1}    
 \end{equation}
 Using conservation of relativistic energy, which amounts to setting the scalar on the right hand side to zero, we have:
 \begin{equation}
 E_{2}^{2}-p_{2}^{2}-m^{2}=0
 \end{equation}
 and
 \begin{equation}
 2m(\omega_{1}-\omega_{2})-2\omega_{1}\omega_{2}(1-\cos\theta)=0    
 \end{equation}
 which results in the familiar expression of Compton \cite{feynman2018theory}:
 
 \begin{equation}
     \dfrac{1}{\omega_{2}}-\dfrac{1}{\omega_{1}}=\dfrac{1}{m}(1-\cos\theta)
 \end{equation}

This is all in accordance with known theory. Let us now discuss the effect of introducing a sinusoidal mass term which varies with some extra parameter, as we have seen with regards to the Majorana particle when viewed through the lens of time optimal quantum state control. We shall also discuss the introduction of a local phase to the momentum vector, and see whether there is any major difference. Computing the anticommutator, we have:
\begin{widetext}
\begin{equation}
\dfrac{1}{2}(\mathbf{p}\mathbf{q}+\mathbf{q}\mathbf{p})=\dfrac{1}{2}\left(\left[\begin{array}{cc}
(\vec{p}\cdot\mathbf{\sigma})(\vec{q}\cdot\mathbf{\sigma}) & 0\\
0 & (\vec{p}\cdot\mathbf{\sigma})(\vec{q}\cdot\mathbf{\sigma})
\end{array}\right]+\left[\begin{array}{cc}
(\vec{q}\cdot\mathbf{\sigma})(\vec{p}\cdot\mathbf{\sigma}) & 0\\
0 & (\vec{q}\cdot\mathbf{\sigma})(\vec{p}\cdot\mathbf{\sigma})
\end{array}\right]\right)    
\end{equation}

Using the identity 
\begin{equation}
(\vec{a}\cdot\mathbf{\sigma})(\vec{b}\cdot\mathbf{\sigma})=(\vec{a}\cdot\vec{b})\mathbf{1}+i(\vec{a}\times\vec{b})\cdot\sigma
\end{equation}
we get:
\begin{equation}
\dfrac{1}{2}(\mathbf{p}\mathbf{q}+\mathbf{q}\mathbf{p})=\dfrac{1}{2}\left(\left[\begin{array}{cc}
(\vec{p}\cdot\vec{q})\mathbf{1}+i(\vec{p}\times\vec{q})\cdot\sigma & 0\\
0 & (\vec{p}\cdot\vec{q})\mathbf{1}+i(\vec{p}\times\vec{q})\cdot\sigma
\end{array}\right]+\left[\begin{array}{cc}
(\vec{q}\cdot\vec{p})\mathbf{1}+i(\vec{q}\times\vec{p})\cdot\sigma & 0\\
0 & (\vec{q}\cdot\vec{p})\mathbf{1}+i(\vec{q}\times\vec{p})\cdot\sigma
\end{array}\right]\right)  
\end{equation}
\begin{equation}
    =(\vec{p}\cdot\vec{q})\mathbf{1}  
\end{equation}
This is for the case of spacelike vectors that share the same phase. If we introduce a phase between the particles in the matrix we may write:
\begin{equation}
\mathbf{p}\mathbf{q}=\left[\begin{array}{cc}
0 & -ie^{-i\theta}\vec{p}\cdot\mathbf{\sigma}\\
ie^{+i\theta}\vec{p}\cdot\mathbf{\sigma} & 0
\end{array}\right]\left[\begin{array}{cc}
0 & -i\vec{q}\cdot\mathbf{\sigma}\\
i\vec{q}\cdot\mathbf{\sigma} & 0
\end{array}\right]=\left[\begin{array}{cc}
(\vec{p}\cdot\mathbf{\sigma})(\vec{q}\cdot\mathbf{\sigma})e^{-i\theta} & 0\\
0 & e^{+i\theta}(\vec{p}\cdot\mathbf{\sigma})(\vec{q}\cdot\mathbf{\sigma})
\end{array}\right]
\end{equation}

and similarly, by symmetry:
\begin{equation}
\mathbf{q}\mathbf{p}=\left[\begin{array}{cc}
(\vec{q}\cdot\mathbf{\sigma})(\vec{p}\cdot\mathbf{\sigma})e^{+i\theta} & 0\\
0 & e^{-i\theta}(\vec{q}\cdot\mathbf{\sigma})(\vec{p}\cdot\mathbf{\sigma})
\end{array}\right]
\end{equation}
In this case, the anticommutator will then take the form:
\begin{equation}
\dfrac{1}{2}(\mathbf{p}\mathbf{q}+\mathbf{q}\mathbf{p})=\dfrac{1}{2}\left[\begin{array}{cc}
(\vec{q}\cdot\mathbf{\sigma})(\vec{p}\cdot\mathbf{\sigma})e^{+i\theta}+(\vec{p}\cdot\mathbf{\sigma})(\vec{q}\cdot\mathbf{\sigma})e^{-i\theta} & 0\\
0 & e^{+i\theta}(\vec{p}\cdot\mathbf{\sigma})(\vec{q}\cdot\mathbf{\sigma})+e^{-i\theta}(\vec{q}\cdot\mathbf{\sigma})(\vec{p}\cdot\mathbf{\sigma})
\end{array}\right]
\end{equation}
\begin{equation}
    =\cos\theta(\vec{p}\cdot\vec{q})\mathbf{1}-\left[\begin{array}{cc}
i\sin\theta(\vec{p}\times\vec{q})\cdot\sigma & 0\\
0 & -i\sin\theta(\vec{p}\times\vec{q})\cdot\sigma
\end{array}\right]
\end{equation}
We can see how the interaction of any two particles is finely balanced in terms of the frequency of interaction. Any difference in this value will result in extra terms for the anticommutation brackets. With regards to the Majorana representation, we calculated a matrix in the first section of the paper:
\begin{equation}
\tilde{H}(t)=i\hat{\beta}m+\hat{\mathbf{\alpha}}\cdot\mathbf{p}=\left[\begin{array}{cccc}
p_{x} & p_{z} & (p_{y}+im)e^{-2iEt} & 0\\
p_{z} & -p_{x} & 0 & (p_{y}+im)e^{-2iEt}\\
(p_{y}-im)e^{2iEt} & 0 & -p_{x} & -p_{z}\\
0 & (p_{y}-im)e^{2iEt} & -p_{z} & p_{x}
\end{array}\right]    
\end{equation}
\end{widetext}
In terms of the anticommutation brackets as discussed in the previous formula, we are interested in the matrix:
\begin{equation}
\mathbf{p}(\phi)=\left[\begin{array}{cccc}
p_{x} & p_{z} & p_{y}e^{-i\phi} & 0\\
p_{z} & -p_{x} & 0 & p_{y}e^{-i\phi}\\
p_{y}e^{i\phi} & 0 & -p_{x} & -p_{z}\\
0 & p_{y}e^{i\phi} & -p_{z} & p_{x}
\end{array}\right]    
\end{equation}
Computing the anticommutation relationship, with one fixed phase, we find:
\begin{equation}
\dfrac{1}{2}(\mathbf{p}\mathbf{q}+\mathbf{q}\mathbf{p})=(p_{x}q_{x}+\cos\phi p_{y}q_{y}+p_{z}q_{z})\mathbf{1}    
\end{equation}
We can see the difference in behaviour here directly. Any variation in phase will again cause a deviation from the expected relationship we expect from the dot product and anticommutation brackets. In this case, a change in phase causes the dot product to effectively become non-positive semi definite. In a directly analogous way, we can set up the scattering problem for the Compton effect:
\begin{equation}
    im\hat{\beta}=\mathbf{p}_{1}
\end{equation}
\begin{equation}
    iE_{2}\hat{\beta}+p_{2}(\hat{\alpha}_{x}\cos\phi+\hat{\alpha}_{y}\sin\phi)=\mathbf{p}_{2}
\end{equation}
\begin{equation}
    \omega_{1}(i\hat{\beta}+\hat{\alpha}_{x})=\mathbf{q}_{1}
\end{equation}
\begin{equation}
    \omega_{2}(i\hat{\beta}+(\hat{\alpha}_{x}\cos\theta+\hat{\alpha}_{y}\sin\theta))=\mathbf{q}_{2}
\end{equation}
Squaring as in the previous calculation, we obtain
\begin{equation}
\mathbf{p}_{2}^{2}=(\mathbf{p}_{1}+\mathbf{q}_{1}-\mathbf{q}_{2})^{2}
\end{equation}
\begin{equation}
\mathbf{p}_{2}^{2}=\mathbf{p}_{1}^{2}+\left\{ (\mathbf{q}_{1}-\mathbf{q}_{2}),\mathbf{p}_{1}\right\} -\left\{ \mathbf{q}_{1},\mathbf{q}_{2}\right\}     
\end{equation}
\begin{equation}
-2m(\omega_{1}-\omega_{2})+2(1-\cos\theta)\omega_{1}\omega_{2}=0
\end{equation}
\begin{equation}
E_{2}^{2}-p_{2}^{2}-m^{2}=0
\end{equation}
\begin{equation}
\dfrac{1}{m}(1-\cos\theta)=\dfrac{1}{\omega_{1}\omega_{2}}\left(\omega_{1}-\omega_{2}\right)=\dfrac{1}{\omega_{2}}-\dfrac{1}{\omega_{1}}
\end{equation}
We can see, to first order, that the Majorana type particle will scatter in the same way as a Dirac electron. However, the difference in the multiplication law will result in a divergence in behaviour for other configurations. We have not considered the effect of a rotating mass in this calculation. We shall discuss in the next section one reason in which this may be acceptable, as it is possible to use transformations between systems of time dependent dynamic operators and fixed states versus static operators and time dependent states to resolve this difficulty.

\section{\label{sec:level1} Different Frames of Reference}
With regard to the basic symmetries of quantum mechanics, it is an established principle that one may use a reference frame whereby all the operators are fixed, and the state changes with the underlying parameter; or the contrary situation where the operator changes with time for fixed state values. We may write this as the equivalence:

\begin{equation}
\left\langle \left.\Psi(t)\left|\tilde{H}(0)\right|\right.\Psi(t)\right\rangle =\left\langle \left.\Psi(0)\left|\tilde{H}(t)\right|\right.\Psi(0)\right\rangle    
\end{equation}

\begin{widetext}

Let us see what consequences this symmetry has for the Hamiltonian of the Majorana particle.

\begin{equation}
    \left|\Psi(t)\right\rangle =\left[\begin{array}{c}
w_{1}\\
w_{2}\\
w_{3}\\
w_{4}
\end{array}\right],
 \left|\Psi(0)\right\rangle =\left[\begin{array}{c}
v_{1}\\
v_{2}\\
v_{3}\\
v_{4}
\end{array}\right]
\end{equation}

\begin{equation}
    \tilde{H}(t)=\left[\begin{array}{cccc}
p_{x} & p_{z} & (p_{y}+im)e^{-2iEt} & 0\\
p_{z} & -p_{x} & 0 & (p_{y}+im)e^{-2iEt}\\
(p_{y}-im)e^{2iEt} & 0 & -p_{x} & -p_{z}\\
0 & (p_{y}-im)e^{2iEt} & -p_{z} & p_{x}
\end{array}\right]
\end{equation}
\begin{equation*}
\left\langle \left.\Psi(t)\left|\tilde{H}(0)\right|\right.\Psi(t)\right\rangle =im(\bar{w}_{1}w_{3}-\bar{w}_{3}w_{1}+\bar{w}_{2}w_{4}-\bar{w}_{4}w_{2})+p_{x}\left(|w_{1}|^{2}-|w_{3}|^{2}+|w_{4}|^{2}-|w_{2}|^{2}\right)
\end{equation*}
\begin{equation}
    +p_{y}\left(\bar{w}_{1}w_{3}+\bar{w}_{3}w_{1}+\bar{w}_{2}w_{4}+\bar{w}_{4}w_{2}\right)+p_{z}\left(\bar{w}_{1}w_{2}+\bar{w}_{2}w_{1}-\bar{w}_{3}w_{4}-\bar{w}_{4}w_{3}\right)
\end{equation}
\begin{equation*}
    \left\langle \left.\Psi(0)\left|\tilde{H}(t)\right|\right.\Psi(0)\right\rangle =im(\bar{v}_{1}v_{3}e^{-2iEt}-\bar{v}_{3}v_{1}e^{2iEt}+\bar{v}_{2}v_{4}e^{-2iEt}-\bar{v}_{4}v_{2}e^{2iEt})+p_{x}\left(|v_{1}|^{2}-|v_{3}|^{2}+|v_{4}|^{2}-|v_{2}|^{2}\right)
\end{equation*}
\begin{equation}
    +p_{y}\left(\bar{v}_{1}v_{3}e^{-2iEt}+\bar{v}_{3}v_{1}e^{2iEt}+\bar{v}_{2}v_{4}e^{-2iEt}+\bar{v}_{4}v_{2}e^{2iEt}\right)+p_{z}\left(\bar{v}_{1}v_{2}+\bar{v}_{2}v_{1}-\bar{v}_{3}v_{4}-\bar{v}_{4}v_{3}\right)
\end{equation}
From this, and in accordance with the result for the diagonalisation procedure, we find the values:
\begin{equation}
    |w_{1}|^{2}-|w_{3}|^{2}+|w_{4}|^{2}-|w_{2}|^{2}=|v_{1}|^{2}-|v_{3}|^{2}+|v_{4}|^{2}-|v_{2}|^{2}
\end{equation}
\begin{equation}
\bar{w}_{1}w_{2}+\bar{w}_{2}w_{1}-\bar{w}_{3}w_{4}-\bar{w}_{4}w_{3}=\bar{v}_{1}v_{2}+\bar{v}_{2}v_{1}-\bar{v}_{3}v_{4}-\bar{v}_{4}v_{3}    
\end{equation}
\begin{equation}
\bar{w}_{1}w_{3}+\bar{w}_{3}w_{1}+\bar{w}_{2}w_{4}+\bar{w}_{4}w_{2}=\bar{v}_{1}v_{3}e^{-2iEt}+\bar{v}_{3}v_{1}e^{2iEt}+\bar{v}_{2}v_{4}e^{-2iEt}+\bar{v}_{4}v_{2}e^{2iEt}
\end{equation}
\begin{equation}
\bar{w}_{1}w_{3}-\bar{w}_{3}w_{1}+\bar{w}_{2}w_{4}-\bar{w}_{4}w_{2}=\bar{v}_{1}v_{3}e^{-2iEt}-\bar{v}_{3}v_{1}e^{2iEt}+\bar{v}_{2}v_{4}e^{-2iEt}-\bar{v}_{4}v_{2}e^{2iEt}
\end{equation}
which results in the unitary relation:
\begin{equation}
    \left[\begin{array}{cccc}
e^{-2iEt} & 0 & 0 & 0\\
0 & e^{-2iEt} & 0 & 0\\
0 & 0 & 1 & 0\\
0 & 0 & 0 & 1
\end{array}\right]\left[\begin{array}{c}
v_{1}\\
v_{2}\\
v_{3}\\
v_{4}
\end{array}\right]=\left[\begin{array}{c}
w_{1}\\
w_{2}\\
w_{3}\\
w_{4}
\end{array}\right]
\end{equation}
we derived earlier from the eigenmatrix of the system.

\end{widetext}
We can see the complex interplay between the various different field mechanisms here, and note that contrary to the Dirac-type system derived in \cite{morrison2019time2}, this system has a mass which is not constant in time. It might be argued that one could equally insert a contrary motion into the state vector, but that would be contradicting our initial assumption, being that the the initial state is fixed in time. One might also attempt to relabel the momentum and mass variables, but that would no longer be the Majorana representation as given.

Interestingly, this gives a natural way in which to associate functions with particular transformation properties under the matrix multiplication law given by the unitary operator. By solving the differential equations for the control fields, we arrived at expressions for the eigenmatrix of states that evolved in time according to the unitary matrix given above. Effectively, by using the symmetry and having access to the expression for the time dependence of the Hamiltonian operator, we can utilise this information to readily write down any form of state evolving forward in time. Conservation laws require that we satisfy certain fundamental relations on vector norms, which we can see expressed as the terms which appear as inner products above. The other terms give different forms of dynamical interactions that must be satisfied; tracing the underlying cause of such relations leads to the anticommutation rules of the spinor representation.

If we examine the squared component of the Hamiltonian matrix, we have immediately that:
\begin{equation}
    \left\langle \left.\Psi(0)\left|\tilde{H}^{2}(t)\right|\right.\Psi(0)\right\rangle =\left\langle \left.\Psi(t)\left|\tilde{H}^{2}(0)\right|\right.\Psi(t)\right\rangle 
\end{equation}
\begin{equation}
    \left\langle \Psi(0)|\Psi(0)\right\rangle \left(|\vec{p}|^{2}+m^{2}\right)=\left\langle \Psi(t)|\Psi(t)\right\rangle \left(|\vec{p}|^{2}+m^{2}\right)
\end{equation}
We can see once more the relationship between the conservation laws and inner products for the time dependent quantum state. This can be derived most easily from the Klein-Gordon relationship
\begin{equation}
    \tilde{H}^{2}(t)=\tilde{H}^{2}(0)=\left(|\vec{p}|^{2}+m^{2}\right)\mathbf{1}
\end{equation} which is obeyed by the spinor matrix.

\section{\label{sec:level1}Discussion and Conclusions}
This paper has covered some new territory in terms of the development of a workable time optimal quantum state control theory. We have shown conclusively that one aspect of existence of physical particles is not necessarily encapsulated in the existing theory. A simple ansatz that the rest mass of any hypothetical particle be a constant when transformed under the unitary laws of motion is perhaps the most straightforward adjustment that can be made. It is interesting to see the direct differences, reflected in the spinor multiplication laws, between the Majorana and Dirac type models. With appropriate adjustments, these systems are feasible ways in which to test the underlying quantum dynamic principles through scattering experiments. Obviously the lack of an observable Majorana particle cannot be directly traced to the failure of its rest mass to transform correctly under the quantum brachistochrone; however, in the absence of absent particles, perhaps it may be possible to find other types of matrix systems with similar properties and show that these also fail to transform and do not exist. 

We have shown also that, to first order, scattering theory will be unable to distinguish Majorana particles from electrons. It is interesting to ponder the result for the same calculation, both for Dirac and Majorana electrons, that would would result if the system has some non-zero phase as expressed through the time dependent Hamiltonian matrix. These types of experiments are the simplest to perform and offer a real chance to prove this theory of time optimal control. Perhaps in the future it may be possible to create synthetic Majorana spinor systems in order to study their properties. It is important to note that while the Majorana system fails the test of analysis, the four dimensional momentum/angular momentum tensor satisfies the quantum brachistochrone and gives us meaningful laws of physics. These may be applied in other contexts, using the process of transformation. From these basic laws of matrix calculus we are able to derive many other dynamic properties of quantum systems, including scattering, unitary operators and many other complex phenomena.

This paper has not addressed the representation theory of the unitary groups we have generated nor other quantum contexts within which the theory of the quantum brachistochrone may be applied. We hope to present further results utilising this theory in the near future, in particular the topics of superconductivity, thermal transfer and further applications within hyperbolic systems.

\begin{acknowledgments}
This research was supported under the UTS Research Excellence Scholarships program, University of Technology, Sydney.
\end{acknowledgments}

\bibliography{apssamp}% Produces the bibliography via BibTeX.

\end{document}